\documentclass[hidelinks,onefignum,onetabnum]{siamart220329}

\usepackage{amsmath}
\usepackage{amssymb}

\newcommand{\rvline}{\hspace*{-\arraycolsep}\vline\hspace*{-\arraycolsep}}

\usepackage{tabularx}
\usepackage{verbatim}
\usepackage{subcaption}
\usepackage{units}

\usepackage{nomencl}
\makenomenclature
\usepackage{makecell}

\usepackage{todonotes}

\usepackage{lipsum}
\usepackage{amsfonts}
\usepackage{graphicx}
\usepackage{epstopdf}

\usepackage{algorithm} 
\usepackage{algpseudocode} 

\usepackage{soul}

\ifpdf
  \DeclareGraphicsExtensions{.eps,.pdf,.png,.jpg}
\else
  \DeclareGraphicsExtensions{.eps}
\fi

\newcommand{\change}[1]{{#1}}

\newcommand{\PT}[1]{\ensuremath{\cdot 10^{#1}}}

\newsiamremark{remark}{Remark}
\newsiamremark{hypothesis}{Hypothesis}
\crefname{hypothesis}{Hypothesis}{Hypotheses}
\newsiamthm{claim}{Claim}

\headers{Graph Random Walk for Time-of-Flight Charge Mobilities}{Zhongquan Chen, Pim van der Hoorn, and Bj\"orn Baumeier}

\title{A Graph Random Walk Method for Calculating Time-of-Flight Charge Mobility in Organic Semiconductors from Multiscale Simulations\thanks{Submitted to the editors DATE.
\funding{B.B. acknowledges support by the Innovational Research Incentives Scheme Vidi of the Netherlands Organisation for Scientific Research (NWO) with project number 723.016.002. We also are grateful for funding and support from ICMS via project MPIPICMS2019001.}}}

\author{Zhongquan Chen\thanks{Department of Mathematics and Computer Science \& Institute for Complex Molecular Systems, Eindhoven University of Technology, PO Box 513, 5600MB Eindhoven, the Netherlands
  (\email{z.chen3@tue.nl}, \email{w.l.f.v.d.hoorn@tue.nl}, \email{b.baumeier@tue.nl}).}
  \and Pim van der Hoorn\footnotemark[2]
\and Bj\"orn Baumeier\footnotemark[2]}

\usepackage{amsopn}

\ifpdf
\hypersetup{
  pdftitle={A Graph Random Walk Method for Calculating Time-of-Flight Charge Mobility in Organic Semiconductors from Multiscale Simulations},
  pdfauthor={Zhongquan Chen, Pim van der Hoorn, and Bj\"orn Baumeier}
}
\fi

\begin{document}

\maketitle

\begin{abstract}
We present a graph random walk (GRW) method for the study of charge transport properties of complex molecular materials in the time-of-flight regime. The molecules forming the material are represented by the vertices of a directed weighted graph, and the charge carriers are random walkers. The edge weights are rates for elementary jumping processes for a charge carrier to move along the edge and are determined from a combination of the energies of the involved vertices and an interaction strength. Exclusions are built into the random walk to account for the Pauli exclusion principle. In time-of-flight experiments, charge carriers are injected into the material and the time until they reach a collecting electrode is recorded. In this setting, our GRW approach allows direct evaluation of the expected hitting time of the collecting nodes in the graph in terms of a typically sparse, linear system, thereby avoiding numerically cumbersome and potentially fluctuations-prone methods based on explicit time evolution from solutions of a high-dimensional system of coupled ordinary differential equations (the Master Equation) or from kinetic Monte Carlo (KMC). We validate the GRW approach by conducting numerical studies of charge dynamics of single and multiple carriers in diffusive and drift-diffusive (due to an external electric field) regimes using a surrogate lattice model of a realistic material whose properties have been simulated within a multiscale model framework combining quantum-mechanical and molecular-mechanics methods. The surrogate model allows varying types and strengths of energetic disorder from the reference baseline. Comparison with results from the Master Equation confirms the theoretical equivalence of both approaches also in numerical implementations. We further show that KMC results show substantial deviations due to inadequate sampling. All in all, we find that the GRW method provides a powerful alternative to the more commonly used methods without sampling issues and with the benefit of making use of sparse matrix methods. 
\end{abstract}



\section{Introduction}
\begin{figure}[tb]
    \centering
    \includegraphics[width=\textwidth]{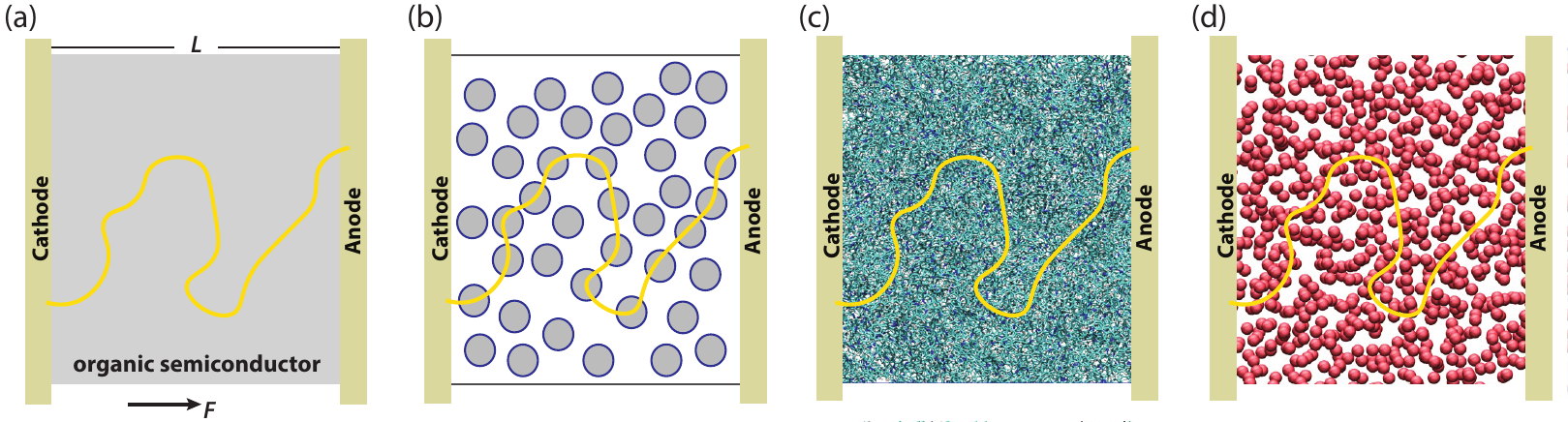}
    \caption{Schematic representations of time-of-flight setups used in experiment and simulations of charge transport: (a) basic setup indicating the electrode setting and the externally applied electric field; (b) circles introduced to emphasize the disordered nature of the molecular material; (c) all-atom molecular detail of an organic semiconductor; (d) center-of-mass view of the molecular material. In all panels, the yellow line indicated an effective charge carrier trajectory.}
    \label{fig:TOF}
\end{figure}

Random walks in random environments are often used to model physical processes~\cite{hughesRandomWalksRandom1996}, where the motion of a particle is represented by the random walker and the random environment is a model for a system that is characterized by some type of disorder, either spatial, temporal, or some other more abstract degree of freedom. Specifically, continuous time random walks (CTRWs) are commonly used in studies of transport processes in materials~\cite{Zeitouni2004}. Organic semiconductors are a particular class of materials that are formed by organic molecules as building blocks which typically arrange in spatially disordered, often amorphous, structures on a mesoscopic scale~\cite{RIEDE2011448}. Such structural disorder translates into a disordered {\em electronic structure}, i.e., quantum states that the electrons can be found in, and both determine the charge transport properties of the material. This is also referred to as {\em energetic disorder}. From the point of view of quantum electronic structure (on the scale of a few \AA) charge transport on a material or device scale (\unit[10-100]{nm}) is an inherently multiscale process~\cite{doi:10.1021/cr040084k}, which lends itself to be modeled as a CTRW. One of the quantities of interest \change{needed} to  characterize charge transport in the presence of an external electric field $\vec{F}$ (e.g., from an applied voltage in an electronic device) is charge mobility~\cite{pasveer_unified_2005}, $\mu = \frac{\vec{v}\cdot\vec{F}}{|\vec{F}|^2}$, where $\vec{v}$ is the average velocity of the charge carrier(s). In practice, this mobility is often obtained from {\em time-of-flight} (ToF) experiments~\cite{tof_exp}, whose setup is sketched in Fig.~\ref{fig:TOF}(a): the organic material is sandwiched by two electrodes, charge carriers are injected from one of these into the material, and are collected at the opposite end. Measuring the time from injection to collection, the time-of-flight $\tau$, and knowing the length of the sample, $L$, allows straightforward calculation of the average velocity along the indicated field direction. Figure~\ref{fig:TOF}(b) is a cartoon emphasizing the disordered nature of the organic semiconductor and the nature of the electronic transport process as a sequence of transfer processes among the molecular building blocks.    

Material design efforts often focus on finding a material with charge mobility that is optimal for a specific purpose. Here, the development of multiscale simulation frameworks that explicitly link the structure of the material at the resolution of atoms to the electronic degrees of freedom by combining quantum and classical first-principles approaches~\cite{ruhle_microscopic_2011} and using them to define the CTRW model in a graph has gained a lot of attention. Figure~\ref{fig:TOF}(c) shows a material at full atomistic resolution, while in Fig.~\ref{fig:TOF}(d), the respective molecular center-of-masses represent the vertices for the graph random walk (GRW), compare to Fig.~\ref{fig:TOF}(b). The edges of the graph can be defined with the help of explicitly calculated bi-molecular transfer rates, as will be presented in Section~\ref{ssec:system_description}. The attractiveness of such frameworks stems from the fact that they allow disentangling the different contributions of physical and chemical properties of the building blocks and their interplay in the full material system, leading to increased understanding at a level typically unattainable by experiments and thereby aiding material design with the definition of particular design rules~\cite{doi:10.1021/ja305310r,stenzel_general_2014,poelkingImpactMesoscaleOrder2015}. Most works on such multiscale models report charge mobilities that are determined in a steady state setting: in contrast to the ToF scenario, no electrodes are taken into account and the material is infinitely repeated in all three dimensions by periodic (cyclic) boundary conditions. Then, the graph dynamics are simulated numerically either by solving a set of ordinary differential equations (also known as Master equation~\cite{omura_master-equation-based_2021} (MEq), see Section~\ref{ssec:master_equation}) explicitly, or by kinetic Monte Carlo~\cite{kolesnikov_kinetic_2018} (KMC, see Section~\ref{ssec:kmc}). Application of these methods can be challenging, e.g., caused by timescale disparities that require special ODE solvers, in how to account for the effect of Pauli exclusion in the case of multiple charge carriers of the same type, due to the trapping of carriers in trap regions, or by sampling issues. However, even if these challenges are met~\cite{stenzel_general_2014,brereton_efficient_2012}, there is no guarantee that such steady-state mobilities are quantitatively, and in some cases even qualitatively, comparable to ToF mobilities.
 
\change{In this paper, we propose using GRW theory to (i) obtain ToF mobilities instead of steady-state mobilities from multiscale simulations and (ii) avoid simulating the explicit dynamics of charge carriers.}
The key aspect of the approach is the calculation of the expected hitting times of random walkers in an absorbing Markov chain. To account for Pauli repulsion of carriers, we construct the associated CTRW with exclusions by a definition of the allowed state configurations of the Markov chain. To demonstrate the capabilities of the GRW method, we use surrogate lattice models as prototypical systems instead of explicit data from multiscale simulations. The basic parameters of the lattice model are chosen based on data from a realistic molecular system, i.e., an amorphous phase of tris(8-hydroxyquinoline)aluminum, an organic semiconductor with high energetic disorder and commonly used in organic light-emitting diodes as an electron-transport material and emitting layer material. Using a lattice model allows straightforward modification of the energetic disorder modeled by Gaussian density-of-states with and without spatial correlations, and tuning between zero and high disorder cases, to thoroughly scrutinize the GRW approach for a variety of different material classes.  

In what follows, we define the system setting, that is the graph and CTRW with the exclusion model for multiple charges, as well as the multiscale and lattice models in Section~\ref{sec:setting}. In Section~\ref{sec:methods} we present the details of the calculation of the hitting times of charge carriers in the CTRW setting from the GRW method and summaries of the alternative  Master equation and kinetic Monte Carlo approaches. Numerical results of the different system settings, including comparisons of GRW to KMC and MEq results, and predictions of electric-field dependent ToF mobilities for different numbers of charge carriers are reported and discussed in Section~\ref{sec:results}. 
In Section~\ref{sec:scale} the scalability of the proposed method is shown and discussed, highlighting its applicability and aspects that need further improvements. 
A brief summary concludes the paper.

\section{System setting}
\label{sec:setting}

We shall model the material as a directed graph $G = (V,E)$, whose vertices are the sites $V = \{1,2,\cdots,n\}$, where some sites are designated as \emph{sources} and some as \emph{sinks}. The directed edges $E$ (often called arcs) between sites are based on the specific choices of our system, see Section~\ref{ssec:system_description} for details. The dynamics of a single charge carrier can then be modeled as a continuous time random walk (CTRW) on this graph with transition rates between connected sites $i$ and $j$ given by some rates $\omega_{ij} > 0$. The directed nature of our graph comes from the fact that we do not impose the rate $\omega_{ij}$ from $i$ to $j$ to be the same as $\omega_{ji}$. The charge dynamic process begins when a charge carrier is injected into one of the source sites and stops when it hits one of the sink sites. 

While this description is for a single charge carrier, the full charge transport dynamics are due to a large number $N_c$ of charge carriers of equal sign moving through the material. Note that this cannot simply be modeled as a collection of $N_c$ independent CTRWs on the graph $G$, because Pauli exclusion implies that each site can only be occupied by no more than one charge carrier. This creates dependencies between the CTRWs of every individual charge carrier. Thus we will model the full dynamics as a CTRW on a larger graph $G_S = (V_S, E_S)$, where each node represents a state describing the position of all charge carriers in the material. 

We will first provide the details for this setup in a general manner, and then describe the elements of the multiscale model and the specific choices made for modeling the charge dynamics in a surrogate lattice model in Section~\ref{ssec:system_description}.

\subsection{A random walk model for multiple charge carriers}
\begin{figure}
    \centering
    \includegraphics[width=0.90\textwidth]{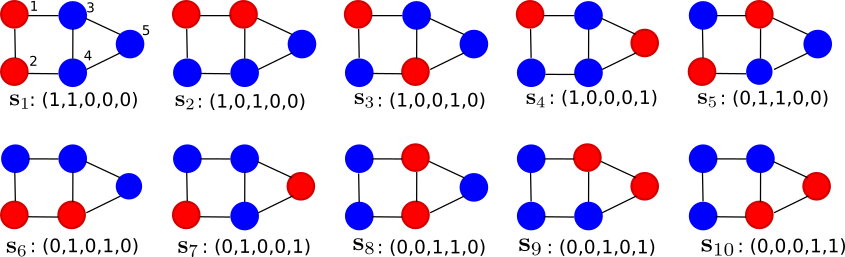}
    \caption{The ten states of a connected 5-site system with blue circles representing unoccupied sites and red circles for occupied sites. The numbers on the top left figure give the site indices. The state notations are the brackets containing 0 and 1 below each state plot. If site $5$ is the sink, then the sink states are $\mathbf{s}_4, \mathbf{s}_7, \mathbf{s}_9, \mathbf{s}_{10}$
    }
    \label{fig:5site}
\end{figure}

The vertex set $V_S$ of our larger graph is the countable ordered set $V_S = \{ \mathbf{s}_1, \mathbf{s}_2, \cdots, \mathbf{s}_{\binom{n}{N_c}} \}$, where each vertex is a configuration of charge \change{carriers} on the sites of the system. More specifically, a vertex $\mathbf{s}$ is a set $\mathbf{s} = ( s_1, s_2, \cdots, s_n)$, whose element $s_i$ is the occupancy of site $i$:
\begin{equation}
  s_i = \begin{cases}
  1 & \text{if}\ i \text{ is Occupied,} \\
  0 & \text{if}\  \text{Unoccupied.}
  \end{cases}
  \label{equ:state_def}
\end{equation}
We will refer to these vertices as \emph{states} and call a state $\mathbf{s}$ a \emph{sink state} when there is at least one site $i$ from the collection of sink sites such that $s_i=1$. \change{A state $\mathbf{s}$ is called a \emph{source state} if all the carrier-occupied sites are source sites. That is, if $s_i=1$, $i$ have to be the source site.}
Figure~\ref{fig:5site} visualizes the 10 states for a 5-site system with two charge carriers.

Next, we define the arcs $E_S$ between the states. There is an arc $(\mathbf{s}, \mathbf{s}')$ if and only if:
\begin{enumerate}
    \item $s_k=s'_k=1$ holds for exactly $N_c-1$ indices $k$, and 
    \item there exists two indices $i,j$ such that $\omega_{ij}> 0$, $s_i=1$ and $s'_i=0$, while $s_j=0$ and $s'_j=1$. 
\end{enumerate}
In this case, we say that $s$ is \emph{connected} to $s^\prime$, while if any of these conditions fail we say that $s$ is \emph{not connected} to $s'$.

Basically, there is an arc from state $\mathbf{s}$ to $\mathbf{s}'$ when starting from the former configuration the latter is obtained by one charge carrier moving over a single arc in the graph $G$ representing the material. For example, in the system shown in Figure~\ref{fig:5site} state $\mathbf{s}_1$ is connected to $\mathbf{s}_3$ and $\mathbf{s}_5$, but is not connected to $\mathbf{s}_2$ or $\mathbf{s}_4$.

For every arc $(\mathbf{s}, \mathbf{s}')$ we can uniquely find two sites $i,j$ such that $\omega_{ij} > 0$, $\mathbf{s}_i =1 = 1-\mathbf{s}'_i$ and $\mathbf{s}'_j =0 = 1-\mathbf{s}_j$. In this case, we will say that this arc is due to $(i,j)$. With this setup we can define the transition rates $\Omega_{s s^\prime}$ between the states in our larger graph as follows: For any two distinct $s \ne s^\prime$
\begin{equation}\label{eq:transition_rates}
	\Omega_{\mathbf{s} \mathbf{s}^\prime} =
	\begin{cases}
	     0			& \text{if } \mathbf{s} \ne \mathbf{s}^\prime \text{ and } \mathbf{s} \text{ is not connected to } \mathbf{s}^\prime,\\
	    \omega_{ij}	& \text{if } \mathbf{s} \ne \mathbf{s}^\prime \text{ and } \mathbf{s} \text{ is connected to } \mathbf{s}^\prime \text{ due to } (i,j), \\
	    -\sum_{\mathbf{t} \ne \mathbf{t}^\prime} \Omega_{\mathbf{t}\mathbf{t}^\prime} & \text{if } \mathbf{s} = \mathbf{s}^\prime.
	\end{cases}
\end{equation}
 
With this setup, the charge transport dynamics correspond to a CTRW on the directed graph $G_S$ with transition rates $\Omega_{\mathbf{s} \mathbf{s}^\prime}$. 

\subsection{The Multiscale Model}
\label{ssec:system_description}
In general, the multiscale model for simulation charge dynamics in organic semiconductors revolves around the calculation (from first principles) of rates of temperature-activated electron transfer between two molecules $i$ and $j$, the transition rates $\omega_{ij}$ (in unit $s^{-1}$) mentioned above. A commonly used expression is the Marcus rate
\begin{equation}
    \omega_{ij} = \frac{2\pi}{\hbar} \frac{|J_{ij}|^2}{\sqrt{4\pi \lambda_{ij} k_\text{B}T}} \exp\left(-\frac{(\Delta\epsilon_{ij} + q \vec{F} \cdot \Vec{r}_{ij} - \lambda_{ij})^2}{4\lambda_{ij} k_\text{B}T}\right) ,
    \label{equ:Marcus}
\end{equation}
where $\hbar$ is the reduced Planck constant, $k_\text{B}$ the Boltzmann constant. The temperature $T$ (in \unit[]{K}), the charge of the carrier $q$ (in \unit[]{e}) and the external electric field $\vec{F}$ (in \unit[]{V/m}) can be considered as parameters of the simulation. The remaining quantities are molecule-pair specific and calculable: the reorganization energy $\lambda_{ij}$, the electronic coupling $J_{ij}$, and the energy difference $\Delta \epsilon_{ij} = \epsilon_i - \epsilon_j$ (all in \unit[]{eV}). The vector $\vec{r}_{ij} = (r^x_{ij},r^y_{ij},r^z_{ij})^\text{T}$ connects the center-of-masses of molecules $i$ and $j$ (the spatial coordinates of vertices $i$ and $j$ in $G$), including cyclic boundary conditions depending on the modeling setting.  Corresponding to the calculable quantities entering \eqref{equ:Marcus}, the multiscale simulation framework comprises three major components: (i) the simulation of the mesoscale material morphology at the resolution of atoms performed with classical Molecular Dynamics from which the vertex and edge sets are defined, (ii) the calculation of sets $\left\{\lambda_{ij}\right\}$,  $\left\{J_{ij}\right\}$, and $\left\{\Delta\epsilon_{ij}\right\}$ based on the atomistic detail from (i) using combinations of quantum electronic structure and classical electrostatics methods, defining the edge weights, and (iii) the evaluation of the charge dynamics with one of the methods mentioned in the Introduction, and to be specified in Section~\ref{sec:methods}. 

In Refs.~\cite{ruhle_microscopic_2011,baumeier_stochastic_2012}, the above framework was applied to a specific organic semiconductor material, an amorphous phase of tris(8-hydroxyquinoline)aluminum (Alq$_3$). Instead of using this model directly, we set up a surrogate model that allows easy definition of different scenarios, specifically for the important energy difference $\Delta \epsilon_{ij}$ in the exponential in \eqref{equ:Marcus}. The surrogate model is based on a regular lattice model in three dimensions, filling the cubic simulation box $[0,L] \times [0,L] \times [0,L] \subset \mathbb{R}^3, L>0$. With lattice constant $a$, and $N$ lattice points per dimension, $L=(N-1)a$, and the position of vertex $i$ is $\vec{r}_i^T=(x_i,y_i,z_i) = (i_x,i_y,i_z)a$ with $i_x,i_y,i_z=0,\ldots,N-1$. 
For the ToF setup, the source and sink regions are located at $x = 0$ and $x = L$, respectively, i.e., we study the time-of-flight along the $x$-direction. While we consider all states involving a site \change{occupied by a charge carrier} with $x=L$ \change{(i.e. $i_x = N-1$)} as the set of sink states, we will specify different choices for the source states from the sites at $x=0$ in Section~\ref{sec:results}. 
\change{As soon as one of the charge carriers reaches the sink regions at $x=L$, the charge dynamics terminates and the ToF is recorded. The sink state can be interpreted as those states containing an occupied site at the sink region.}
\change{With this setup, the connection vector is denoted as $\vec{r}_{ij}^T = \vec{r}_i^T-\vec{r}_j^T = (x_{ij},y_{ij},z_{ij})$, where $x_{ij}=x_i - x_j$.}
For the perpendicular directions $y$ and $z$, we assume cyclic boundary conditions, i.e., for the $y$-component of the vector connecting $i$ and $j$
\begin{equation}
  y_{ij} = \begin{cases}
        y_j - y_i &\text{ if } |y_j - y_i| < L/2, \\
        L + (y_j - y_i) &\text{ if } |y_j - y_i| \geq L/2 \text{ and } y_i > y_j \\
        -L + (y_j - y_i) &\text{ if } |y_j - y_i| \geq L/2 \text{ and } y_i < y_j
    \end{cases}
    \label{equ:period}
\end{equation}
and $z_{ij}$ are defined analogously. We use the so defined entries of the connection vector $\vec{r}_{ij}$ to determine the distance metric $|\vec{r}_{ij}| = \sqrt{ (x_{ij})^2 + (y_{ij})^2 + (z_{ij})^2 }$ and an edge is assigned to all vertex pairs $(i,j)$ for which $|\vec{r}_{ij}| <2a$. For simplicity, we set $a=\unit[1]{nm}$.
\change{In the current setup of the lattice model, the total number of sites $n$ is related to number of lattice points $N$ by $n = N^3$. To ensure computational feasibility for both the Master equation, KMC, and GRW methods, we select $N = 8$. This choice balances the requirements of numerical implementation across all methods for the purpose of comparison and aligns with typical molecular system sizes used in first-principles multiscale modeling.}

As in the explicit multiscale model, we use a single reorganization energy for all pairs, i.e., $\lambda_{ij} = \lambda = \unit[0.23]{eV}$. We model the coupling elements mimicking its known exponential distance dependence by $\vert J_{ij}(|\vec{r}_{ij}|)|^2=J_0\exp{\left(-(|\vec{r}_{ij}|-a)\right)}$ with $J_0 =\unit[1.79\cdot 10^{-4}]{(eV)^2}$. In the multiscale model, the site energies $\epsilon_i$ follow a Gaussian distribution with mean $\Bar{\epsilon}$ and variance $\sigma^2$. In the first class of site-energy models, the $\epsilon_i$ are taken as independent and identically distributed (iid) Gaussian samples. We refer to this model as {\em uncorrelated}. However, in Alq$_3$ (and several other materials) the site energies show spatial correlation. In a {\em correlated} site-energy model, we adopt a moving average procedure as in~\cite{baumeier_stochastic_2012}, making use of the invariance properties of the normal distribution with respect to convolution. First, with three sequences of iid random variables $M_i^{(a)},M_i^{(b)},M_i^{(c)} \sim \mathcal{N}(0,\sigma^2)$, every site is assigned the 4-tuple $(V_i, M_i^{(a)},M_i^{(b)},M_i^{(c)})$ to which we want to allocate a random site energy $\epsilon_i$. If $V_i^{(1)}$,$V_i^{(2)}$,$\cdots$,$V_i^{(l)}$ are the $l$ nearest neighbors of $V_i$ with corresponding random variables $M_i^{(b),(j)},M_i^{(c),(j)}$, $j = 1,2,\cdots,l$ the spatially correlated energies are evaluated as:
\begin{equation}
    \epsilon_i = \sqrt{\eta_a} M_i^{(a)} + \sqrt{\frac{\eta_b}{l_b}} \sum\limits_{j=1}^{l_b} M_i^{(b),(j)} + \sqrt{\frac{1-\eta_a - \eta_b}{l_c}} \sum\limits_{j=1}^{l_c} M_i^{(c),(j)} + \Bar{\epsilon},
    \label{equ:correlation}
\end{equation}
where $\eta_a,\eta_b \geq 0$ ($\eta_a+\eta_b \leq 1$) are the weights for the individual components and $l_b,l_c > 0$ for some integers. Equation \eqref{equ:correlation} develops the spatially correlated $\epsilon_i$ as a superposition of three independent energy landscapes. From the explicit Alq$_3$ data we choose $\Bar{\epsilon}=\unit[-0.76]{eV}$, $\sigma=\unit[0.19]{eV}$, $\eta_a=0.2$, $\eta_b=0.4$, as well $l_b=9$ and $l_c=280$. 

\section{Methods}
\label{sec:methods}
The main physical quantity of interest that we want to study is the mobility (or equivalently time-of-flight) in the system representing the molecular material. Both of these come down to the first time a charge carrier hits one of the sink sites. In terms of our random walk description, this is called the \emph{hitting time} of the sink states. We will first describe our proposed method for computing the hitting time, using our CTRW setup. After this, we compare it to two other known methods: Master Equation and KMC.

\subsection{Hitting time for CTRW system}
Let $\vec{\tau} \in \mathbb{R}_+^{\binom{n}{N_c}}$ be the vector of expected hitting times, where $\tau_\mathbf{s}$ denotes the expected hitting time of \change{sink} states starting from state $\mathbf{s}$. Recall the transition rates $\Omega_{\mathbf{s} \mathbf{s}^\prime}$ from state $\mathbf{s}$ to state $\mathbf{s}^\prime$, define $D_\mathbf{s} := \sum_{\mathbf{s}^\prime \ne \mathbf{s}} \Omega_{\mathbf{s} \mathbf{s}^\prime}$ as the total rate out of state $\mathbf{s}$ and let $p_{\mathbf{s} \mathbf{s}^\prime} = \Omega_{\mathbf{s} \mathbf{s}^\prime}/D_\mathbf{s}$. The latter can be seen as the probability of our charge carrier system going from state $s$ to $s^\prime$. Then the vector $\vec{\tau}$ is defined as the minimal non-negative solution to the following recursive equation, see for example Theorem 3.3.3 in~\cite{norris_markov_1998}

\begin{equation}\label{eq:hitting_time}
	\tau_\mathbf{s} = \begin{cases}
		\frac{1}{D_\mathbf{s}} + \sum_{\mathbf{s}^\prime \ne \mathbf{s}} p_{\mathbf{s} \mathbf{s}^\prime} \tau_{\mathbf{s}^\prime} &\text{if $\mathbf{s}$ is not a sink state},\\
		0 &\text{else.} 
	\end{cases}
\end{equation}

Equation \eqref{eq:hitting_time} can be written in the matrix form by taking the transition matrix $\mathbf{P} \in [0,1]^{\binom{n}{N_c} \times \binom{n}{N_c}}$, whose entries are 
\[
	P_{\mathbf{s}\mathbf{s}^\prime} = \begin{cases}
		0 &\text{if $\mathbf{s} \ne \mathbf{s}^\prime$ and $\mathbf{s}$ is a sink state,} \\
		1 &\text{if $\mathbf{s} = \mathbf{s}^\prime$ is a sink state, and}\\
		p_{\mathbf{s}\mathbf{s}^\prime} &\text{else.}
	\end{cases}
\]
By separating the sink states and the non-absorbing states, the matrix $\mathbf{P}$ can be rearranged into a block matrix:
\begin{equation}\label{equ:matrix_P}
\mathbf{P} = \begin{pmatrix}
  \mathbf{P}_1
  & \rvline & \mathbf{P}_2 \\
\hline
 \mathbf{0} & \rvline &
  \mathbf{I}
\end{pmatrix}
\end{equation}
Here $\mathbf{I}$ is the identity matrix, $\mathbf{P}_1$ the submatrix restricted to the non-sink states, and $\mathbf{P}_2$ is the submatrix of the probability of transition from non-sink vertexes to sink vertexes. Due to the Perron-Frobenius Theorem, the maximal eigenvalue of $\mathbf{P}$ is 1, and hence the matrix $(\mathbf{I}-\mathbf{P})$ is not invertible. However, $(\mathbf{I}-\mathbf{P}_1)$ is invertible. Now, let $\vec{\tau}^\ast$ be the part of the hitting time vector for non-sink states, then this satisfies
\begin{equation}
     (\mathbf{I} - \mathbf{P}_1) \vec{\tau}^\ast = \vec{\tau}^0,
     \label{equ:matrix1}
\end{equation}
where $\tau^0_\mathbf{s} = \frac{1}{D_{\mathbf{s}}}$.  In the context of CTRWs, $\tau^0_\mathbf{s}$ represents the holding time at state $\mathbf{s}$. The role of the matrix $\mathbf{I} - \mathbf{P}_1$ is to transform these holding times into hitting times. To obtain the ToF, we need to consider the elements $\vec{\tau}_\mathbf{s}^\ast$ where $\mathbf{s}$ is a source state. The process of hitting the sink states from individual source states is somewhat reminiscent of a parallel electric network of capacitors~\cite{doyle_random_2000}, and accordingly, we evaluate 
\begin{equation}
\tau = N_\text{source} \left[\sum_{\mathbf{s}\in \text{source}} (\tau_\mathbf{s}^\ast)^{-1}\right]^{-1},
\end{equation}
with $N_\text{source}$ the number of source states, i.e., as the harmonic mean of the respective $\tau_\mathbf{s}^\ast$.

In practical terms, the sparsity of $\mathbf{P}_1$ allows for efficiently solving equation~\eqref{equ:matrix1} using iterative techniques. In particular, we first apply the quasi-minimal residual method~\cite{freund_qmr_1991} to solve equation~\eqref{equ:matrix1} using as an initial guess the all zero vector. The iterative process is considered to be converged when $|| \vec{\tau}^0 - (\mathbf{I} - \mathbf{P}_1) \vec{\tau}^\ast ||_2 \leq \varepsilon \times || \vec{\tau}^0 ||_2 $, where $\varepsilon=10^{-5}$. 
When the iterative steps exceed $10^4$ and this criterion is not yet achieved, the biconjugate gradients stabilized method~\cite{van_der_vorst_bi} is used to solve equation~\eqref{equ:matrix1} using the outcome $\vec{\tau}^\ast$ of the previous method as the initial guess. The convergence criterion is the same.

\subsection{Hitting time using the Master Equation}
\label{ssec:master_equation}
The Master Equation (MEq) method serves as a valuable tool for investigating charge transport phenomena within disordered organic materials~\cite{pasveer_unified_2005, pasveer_scaling_2005, casalegno_numerical_2013, koster_charge_2010}. These investigations typically consider the so-called Pauli master equation for the $P_{\mathbf{s}}(t)$, the probability that the system is state $\mathbf{s}$ at time $t$. For our system, this equation would be
\begin{equation}
    \frac{d P_{ \mathbf{s} }}{d t} = \sum_{\mathbf{s'} } P_{ \mathbf{s}^\prime }(t) \Omega_{\mathbf{s}\mathbf{s}^\prime} - \sum_{\mathbf{s}^\prime } P_{ \mathbf{s} }(t) \Omega_{\mathbf{s}^\prime\mathbf{s}},
    \label{eq:master_equation}
\end{equation}
or in more compact form $\frac{d\vec{P}(t)}{dt} = \boldsymbol{\Omega}\vec{P}(t)$.
This equation is often solved using numerical techniques, e.g.~\cite{yu_molecular_2001}. Given $P_{\mathbf{s}}(t)$, the expected hitting time is easily computed using
\begin{equation}
    \tau = \sum_{\mathbf{s} \in \text{sink}}\int_0^{\infty} t \frac{dP_{\mathbf{s}}(t)}{dt} dt,
    \label{eq:hitting_time_master_eq}
\end{equation}
where $\text{sink}$ denotes the set of sink states and derivatives are taken component-wise. However, as is evident from~\eqref{eq:hitting_time}, our approach does not require knowing $P_{\mathbf{s}}(t)$, which describes the full dynamics. Moreover, both methods yield the same result, as the CTRW described by the transition rates~\eqref{eq:transition_rates} yield the same $P_{\mathbf{s}}(t)$ as the MEq method, see for example Theorem 2.8.2 in \cite{norris_markov_1998}. Hence, our approach computes the required quantity directly, circumventing additional computation challenges \cite{jahnke_solving_2008,wolf_solving_2010} associated with solving~\eqref{eq:master_equation}. 

\subsection{Hitting time using KMC}
\label{ssec:kmc}
An alternative to solving the master equation for studying transport processes in semiconductors and organic cells is Kinetic Monte Carlo (KMC)~\cite{cottaar_modeling_2012, casalegno_numerical_2013, marsh_microscopic_2007, wolf_current_1999, casalegno_coarse-grained_2012}. KMC simulates the dynamic process and calculates the properties of interest. This offers a great advantage, as it can be used to model the behavior of systems of arbitrary complexity providing that the rules governing the processes are known. However, it often requires long computations for each simulation, with a single simulation sometimes requiring hours or even days to calculate the quantities of interest. Furthermore, data distribution limits further analysis, such as parameter fitting to experimental data or conducting sensitivity analyses. Moreover, many simulations need to be conducted to ensure that the results indeed correspond to the true average values of the stochastic model. For an example of these, see Section~\ref{ssec:GRW_vs_KMC} and Figure~\ref{fig:t_step}. Overall, the computational time and convergence for KMC remain the more challenging aspects of the method. 

We specifically use the KMC implementation for multiple charge carriers as in~\cite{ruhle_microscopic_2011}, which employs a two-step approach: first, one carrier is selected from the list of all carriers based on the smallest randomly calculated waiting time using the \emph{First Reaction Method} (FRM)~\cite{FRM}. Then, the destination site of the selected carrier is determined (with occupied destination sites removed due to Pauli exclusion) and the carrier is moved with the \emph{Variable Step Size Method}. Whenever the sink is reached by one carrier, we report the associated system time as $\tau$. In contrast to KMC, our method provides a direct formula for the expected hitting time from the system description. From this, the hitting time can be computed (solving the equations numerically) without the need to run large-scale simulations.

\section{Results}
\label{sec:results}
In this section, we show the results for the time-of-flight $\tau$ and the dependence of the mobility $\mu$ on the strength of the externally applied electric field $\vec{F}$ from the GRW method for different settings and comparisons to MEq and KMC methods. The base model we use is a lattice model as introduced in Section~\ref{ssec:system_description} with $N=8$. \change{This system size is typical for medium-scale molecular systems in multiscale modeling studies (examples can be found in \cite{ruhle_microscopic_2011,cheung_modelling_2008,sato_multiscale_2019,brereton_efficient_2014}). It also ensures that the numerical solution of the master equation, Equation~\eqref{equ:ODE}, remains computationally feasible, enabling a direct comparison between our proposed method and the true dynamics of Equation~\eqref{eq:master_equation}. } 
While based on properties of the explicit multiscale model for Alq$_3$ from~\cite{ruhle_microscopic_2011,baumeier_stochastic_2012}, the choice of a lattice model allows studying different settings for the site energies on an equal footing: the no-disorder case ($\epsilon_i=0$); uncorrelated, Gaussian distributed disorder; and spatially correlated Gaussian disorder as modeled by~\eqref{equ:correlation}. To study different \emph{amounts} of disorder, we introduce a scaling parameter $k\in[0,1]$ into the site energy term in the exponential of the Marcus rate, i.e., we let $\Delta\epsilon_{ij} \to k\Delta\epsilon_{ij}$ in~\eqref{equ:Marcus}, effectively smoothly interpolating from $\sigma=\unit[0.0]{eV}$ to $\sigma=\unit[0.19]{eV}$ using the same reference random samples of the site energies. Further, we consider the case of $N_c=1$ and $N_c=2$ charge carriers.

\subsection{Comparison of Master equation and Graph Random Walk}
In Section~\ref{ssec:master_equation} we have seen that the MEq and GRW approaches provide formally equivalent predictions of expected hitting times for the defined sink regions and with that of the charge carriers' time-of-flight. Here, we aim to validate our numerical implementation of the GRW by comparing its predictions of $\tau$ to those obtained from the MEq. We employ the analytic form of the solution of the initial value problem in~\eqref{eq:master_equation} with the eigenvalues $\{\xi_\alpha\}$ and corresponding eigenvectors $\{\vec{u}_\alpha\}$ of the matrix $\boldsymbol{\Omega}$ as
\begin{equation}
\vec{P}(t)=\sum_{\alpha=1}^{\binom{n}{N_c}} c_\alpha \vec{u}_\alpha e^{\xi_\alpha t},
\label{equ:ODE}
\end{equation}
where $\vec{c} = \mathbf{U}^{-1}\vec{P}(0)$ and $\mathbf{U} = (\vec{u}_1,\ldots,\vec{u}_{\binom{n}{N_c}})$.  As initial value for \eqref{eq:master_equation}, we set $\vec{P}(0)=\begin{cases}N_c/N_\text{source} & \text{for $\mathbf{s}\in$\,source }\\0&\text{else}\end{cases}$, where $N_\text{source}$ is the number of source states defined. The form of~\eqref{equ:ODE} allows analytic evaluation of the integral in~\eqref{eq:hitting_time_master_eq} to obtain $\tau$. 

\begin{figure}[tb]
    \centering
    \includegraphics[width=1.0\linewidth]{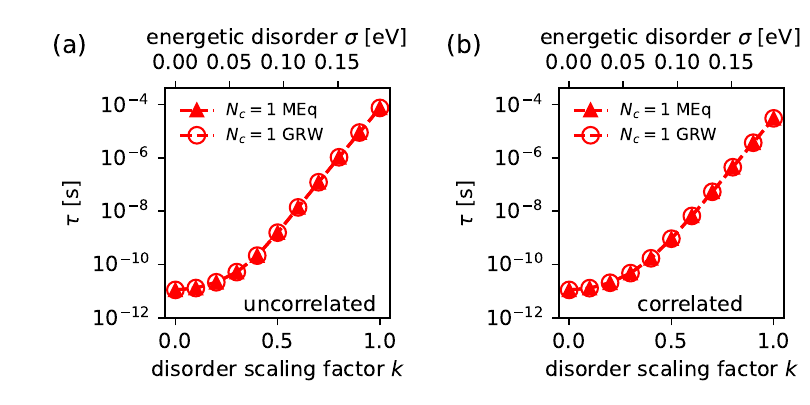}
    \caption{Calculated ToF $\tau$ (in \unit[]{s}) for $N_c=1$ depending on disorder strength in systems with uncorrelated (a) and spatially correlated (b) site energies, obtained from MEq (filled symbols) and GRW (open symbols), respectively. Each data point $\tau$ represents the sample average of the ten realizations of the Gaussian distributed energy landscapes.} 
    \label{fig:RW_eigen}
\end{figure}

For the purpose of comparing the two methods, we consider a single source state for both $N_c=1$ ($\mathbf{s}=(1,0,\ldots,0)$) and $N_c=2$ ($\mathbf{s}=(1,1,0,\ldots,0)$), respectively. Specifically, the site $s_1$ is located at $\vec{r}_1^T=(0,0,0)$ and $s_2$ at $\vec{r}_2^T=(0,0,a)$. In Figure~\ref{fig:RW_eigen} we show the results of the obtained dependence of $\tau$ on $k$ for the systems with uncorrelated and correlated disorder and $N_c=1$, respectively. No external electric field is applied, i.e., $\vec{F}=\vec{0}$, so the dynamics of the charge carriers is purely diffusive. For both MEq and GRW methods, the hitting time $\tau$ is calculated by averaging ten systems each with the same $\Bar{\epsilon},\sigma$. It is visually clear that there is numerically excellent agreement between both approaches, as was expected from the theoretical remarks in Section~\ref{ssec:master_equation}. The relative errors are consistently smaller than $\unit[0.1]{\%}$. 

This comparison confirms the validity of the GRW method we propose, at least for $N_c=1$. As convenient as the form in~\eqref{equ:ODE} is at first glance, its application to larger-scale problems is problematic numerically: it requires the full eigendecomposition of $\boldsymbol{\Omega}$, which cannot be obtained with sparse matrix methods. In the present study, the $N_c=2$ case with $n=512$ requires numerical diagonalization of a full matrix with dimension 130816, even though $\boldsymbol{\Omega}$ is extremely sparse. This highlights one of the advantages of the GRW method as it can make use of efficient sparse matrix implementations throughout. Alternative approaches to solve~\eqref{eq:master_equation} that make use of numerical discretization schemes like forward and backward Euler methods or higher-order Runge--Kutta methods, could be formulated in sparse forms. However, the details of the matrix $\Omega$ make time-stepping methods inefficient. For instance, in one realization of the $k=1$ uncorrelated disorder case, the non-zero eigenvalues of $\Omega$ range from $\unit[-10^{13}]{}$ to $\unit[-10^{2}]{}$. With the associated time-scales of the dynamic modes ranging from $\unit[10^{-13}]{s}$ to $\unit[10^{-2}]{s}$, and the average hitting time of $\unit[\sim 10^{-3}]{s}$, a very small time step ($\unit[10^{-14}]{s}$) and many such steps ($\unit[10^{11}]{}$) might be required in an extreme case to resolve $\tau$ reliably. In addition, the discretization scheme must also allow for integration of~\eqref{eq:hitting_time_master_eq}. All in all, using time-stepping methods that resolve the actual dynamics of the probabilities to obtain $\tau$ seems cumbersome. 

\subsection{Comparison of GRW and KMC}\label{ssec:GRW_vs_KMC} 

\begin{figure}[tb]
    \centering
    \includegraphics[width=1.0\textwidth]{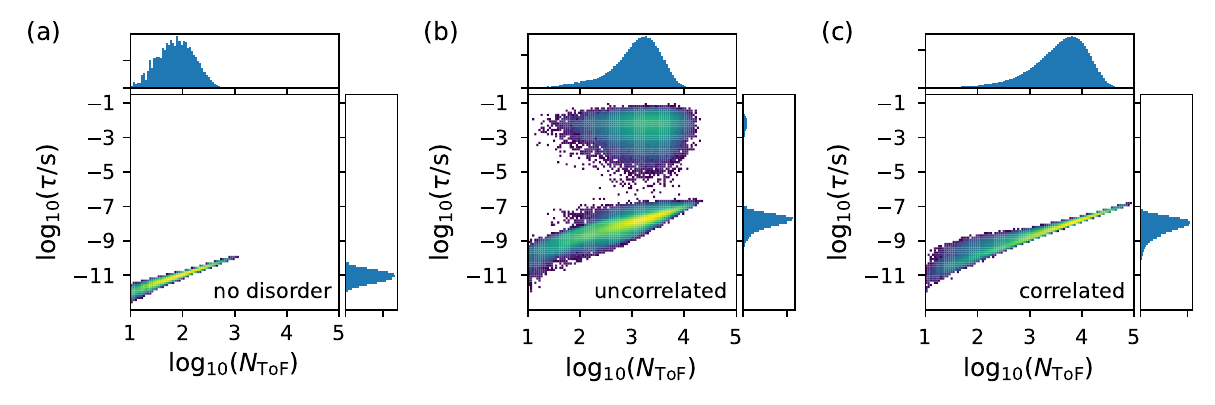}
    \caption{Distributions of KMC steps until absorption to sink ($N_\text{ToF}$) and the corresponding first hitting time $\tau$ obtained via KMC for a single realization of the lattice model with no site-energy disorder (a), Gaussian-distributed site-energies without (b) and with spatial correlations (c). The density histogram consists of $10^7$ KMC sample points grouped into 100 bins.}
    \label{fig:t_step}
\end{figure}

We begin the discussion of the results by highlighting the potentially problematic convergence behavior of KMC simulations as an alternative to solving the explicit dynamics of~\eqref{eq:master_equation}. For a single realization of the site energies with no, uncorrelated, and spatially correlated disorder with $\sigma=\unit[0.19]{eV}$, respectively, we performed $10^7$ KMC  simulations. We recorded for each of these simulations the individual first hitting time of the sink region ($\tau$) and the associated number of KMC steps ($N_\text{ToF}$). The resulting distributions are depicted in Figure~\ref{fig:t_step}. One can see that the no disorder case in panel (a) is largely unremarkable. There is very little noticeable spread in the distributions of the time-of-flight and number of steps recorded, with the maximum number of steps needed being around $1000$. In contrast, the data for the uncorrelated disorder case shown in Figure~\ref{fig:t_step}(b), illustrate one of the key challenges for KMC: the majority of $\tau$ values cluster around $\unit[10^{-8}]{s}$, while only a small fraction falls near $\unit[10^{-2}]{s}$, which is the main contribution to the average hitting time ($\unit[4.3 \cdot 10^{-4}]{s}$). The comparatively rare occurrences of long ToFs can be attributed to the presence of isolated sites with low energies that are not always visited. Most of the KMC simulations seem to require 10000 steps or fewer. In spatially correlated disorder (Figure~\ref{fig:t_step}(c)), we do not observe the influence of apparently rarely sampled sites. Instead, due to the spatial correlation, low-energy sites are not isolated, and several sites are likely to form \emph{regions} of relatively low energy. As the relative energy difference among the sites inside these regions is small, the random walker in the KMC simulations spends a significant number of steps in these regions before it progresses to another of such regions or the sink. Consequently, some KMC simulations require up to $10^5$ steps to finish. Overall, it is clear that depending on the characteristics of the energetic disorder in the material, the results of ToF simulations from KMC simulation might be subject to significant fluctuations, with convergence requiring numerous samples, and in general potentially long simulation times per sample. 

\begin{figure}[tb]
    \centering
    \includegraphics[width=\linewidth]{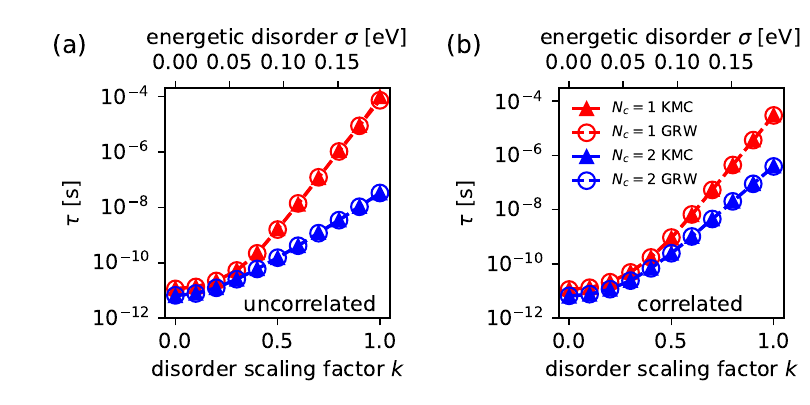}
    \caption{Calculated ToF $\tau$ (in \unit[]{s}) depending on disorder strength on systems with uncorrelated (a) and spatially correlated (b) site energies, obtained from KMC (filled symbols) and GRW (open symbols), respectively. Data points in red (blue) show the results for $N_c=1$ ($N_c=2$). Each data point $\tau$ represents the sample average of the ten realizations of the Gaussian distributed energy landscapes. Each KMC simulation contains 1000 runs.}
    \label{fig:t_RW_KMC}
\end{figure}

Against this background, combined with the difficulties discussed in relation to the MEq in the previous section, the GRW method is expected to provide a significant advantage in not being affected by these and related convergence problems. We demonstrate the quality of the predictions of $\tau$ obtained by solving \eqref{equ:matrix1} compared to a reference from extensive KMC calculations. As in the previous section, the hitting time $\tau$ is calculated by averaging ten systems each with the same $\Bar{\epsilon},\sigma$, with the tuned amount of disorder. The full set of results are listed in Table~\ref{tab:compare} in the Appendix. Figure~\ref{fig:t_RW_KMC} shows the obtained dependence of $\tau$ on $k$ from the two methods for uncorrelated and correlated disorder and for one and two charge carriers, respectively. Again, no external electric field is applied, i.e., $\vec{F}=\vec{0}$.

In the case of KMC, 1000 runs are performed for each realization of the site energies, with carriers populating the source state. Overall there is a very good agreement between the results obtained with the GRW method as compared to KMC. Qualitatively, the data indicate that the more disorder is in the system, the larger the time-of-flight becomes. We also observe in all settings a reduction of $\tau$ as the number of carriers increases, and that $\tau$ decreases more in systems with uncorrelated site energies as compared to those with correlated site energies. This observation aligns with the findings reported in \cite{bouhassoune_carrier_2009}. According to \cite{baranovskii_theoretical_2014}, one contributing factor to this phenomenon is the nature of the isolated low-energy sites in the uncorrelated case vs the possible existence of regions (or small clusters) of low-energy sites in the spatially correlated case. In the former, a single carrier can be trapped in such an isolated low-energy site, and for $N_c=2$, due to the exclusion, the second carrier can ''freely'' diffuse within the rest of the system. This would correspond in Figure~\ref{fig:t_step}(b) to the elimination of the KMC trajectories with the large $\tau$ as discussed above. 

\begin{figure}[tb]
    \centering
    \includegraphics[width=\linewidth]{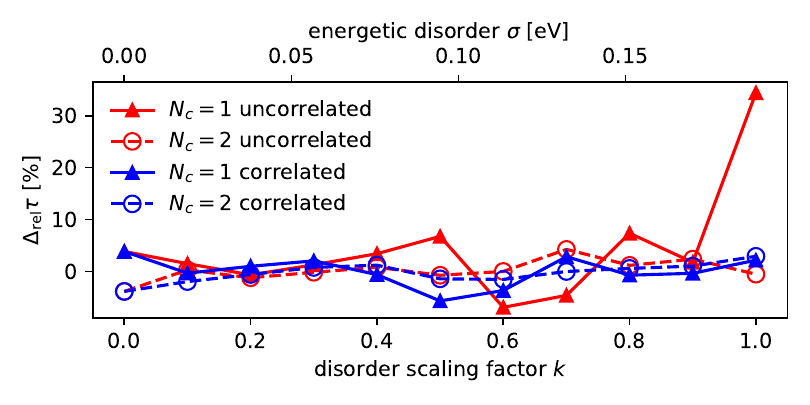}
    \caption{Relative difference $\Delta_\text{rel}\tau = (\tau_\text{KMC} - \tau_\text{GRW})/\tau_\text{GRW}$ (in \unit[]{\%}) for different disorder scaling factors $k$ in systems with uncorrelated and spatially correlated disorder and different numbers of charge carriers, respectively.}
    \label{fig:t_RW_KMC_error}
\end{figure}

From the logarithmic scale on the $y$-axes in Figure~\ref{fig:t_RW_KMC}, it is difficult to assess the differences in the results from GRW and KMC in detail. We therefore show in Figure~\ref{fig:t_RW_KMC_error} the relative difference $\Delta_\text{rel}\tau = (\tau_\text{KMC} - \tau_\text{GRW})/\tau_\text{GRW}$ depending on the disorder scaling factor $k$, for all four cases studied. Except for the case $N_c=1$ in uncorrelated disorder with $k=1$, the results differ at most by $\unit[7]{\%}$ ($N_c=1$ uncorrelated), $\unit[4]{\%}$ ($N_c=2$ uncorrelated),  $\unit[6]{\%}$ ($N_c=1$ correlated), and  $\unit[4]{\%}$ ($N_c=2$ correlated), respectively. 
\change{For the case $N_c=1$ with uncorrelated disorder and $k=1$, a $\unit[30]{\%}$ relative error arise due to the inadequate KMC samplings. This error highlights the convergence limitations of KMC, attributed to insufficient sampling due to the low-energy regions of the molecular system. } 

\subsection{Electric field dependent time-of-flight mobility from GRW}

\begin{figure}[tb]
    \centering
    \includegraphics[width=\linewidth]{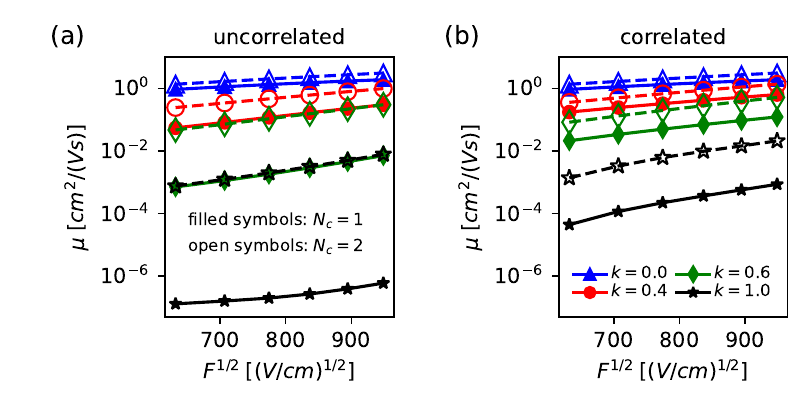}
    \caption{Electric-field dependence of the mobility $\mu$ in the $N=8$ multiscale model system for (a) spatially uncorrelated and (b) spatially correlated disorder for $N_c=1$ (filled symbols) and $N_c=2$ (open symbols) and varying strength of energetic disorder via scaling factor $k$, as obtained from the GRW method. }
    \label{fig:time_F2}
\end{figure}

With the advantages of the GRW method over MEq and KMC approaches as discussed in the previous two sections for the case of purely diffusive charge carrier dynamics, we now turn to the application of the GRW framework to study the electric-field dependent mobility of one and two charge carriers, whose dynamics now correspond to a drift-diffusion process. Recall from the Introduction that the mobility is defined as $\mu = \frac{\vec{v}\cdot\vec{F}}{|\vec{F}|^2}$. In the following, we set $\vec{F}^T=(F,0,0)$, such that in the setting described in Section~\ref{ssec:system_description} we can evaluate the mobility as a function of $F$ as $\mu(F)=\frac{L}{\tau(F) F}$. We use $\tau(F)$ here to emphasize that the ToF itself also depends on the value of the electric field, through the Marcus rates~\eqref{equ:Marcus}. Of course, as seen before for the diffusive carrier dynamics, the ToF also depends on the type and amount of disorder in the system. 

From experiments, it is empirically known that the charge carrier mobility of many disordered organic semiconductors is approximately $\mu(F)=\mu_0\exp(\beta\sqrt{F})$, for $F \in \left[\unit[10^7]{V/m}, \unit[10^8]{V/m} \right]$.  Poole and Frenkel also predicted this electric-field dependence in a model describing the mechanism of trap-assisted electron transport for insulators and semiconductors~\cite{PhysRev.54.647}. It is therefore common to plot the mobility $\mu$ against $\sqrt{F}$ in a so-called \emph{Poole--Frenkel plot}.

\begin{table}[tb]
    \caption{Poole--Frenkel parameters $\mu_0$ (in $\unit[]{cm^2/(Vs)}$) and $\beta$ (in $\unit[]{\sqrt{cm/V}}$) extracted from GRW simulations for $N_c=1$ and $N_c=2$.}\label{tab:PFdata}
    \begin{center}
      \begin{tabular}{cccccc} \hline
        &  \multicolumn{2}{c}{\bf uncorrelated} & &\multicolumn{2}{c}{\bf correlated}\\\cline{2-3}  \cline{5-6}
        $k$  & $\mu_0$ & $\beta$ && $\mu_0$ & $\beta$\\\hline
        \multicolumn{6}{c}{\bf GRW for $\mathbf{N_c=1}$}\\
    0.0  & 2.4\PT{-1} & 2.2\PT{-4} && 2.4\PT{-1} & 2.2\PT{-4}\\
    0.4  & 1.9\PT{-3} & 5.3\PT{-4} && 1.5\PT{-2} & 3.9\PT{-4}\\ 
    0.6  & 2.5\PT{-4} & 7.8\PT{-4} && 8.1\PT{-4} & 5.3\PT{-4}\\
    1.0  & 2.2\PT{-9} & 5.8\PT{-4} && 4.5\PT{-7} & 8.0\PT{-4}\\[0.3cm]
    \multicolumn{6}{c}{\bf GRW for $\mathbf{N_c=2}$}\\
    0.0  & 2.7\PT{-1} & 2.6\PT{-4} && 2.7\PT{-1} & 2.6\PT{-4}\\
    0.4  & 1.6\PT{-2} & 4.4\PT{-4} && 2.7\PT{-2} & 4.1\PT{-4}\\ 
    0.6  & 1.1\PT{-3} & 5.9\PT{-4} && 2.5\PT{-3} & 5.6\PT{-4}\\
    1.0  & 4.6\PT{-6} & 7.8\PT{-4} && 2.1\PT{-5} & 7.3\PT{-4}\\     
    \hline
      \end{tabular}
    \end{center}
\end{table}

We show in Figure~\ref{fig:time_F2} such a plot as resulting from GRW calculations for our system with (a) spatially uncorrelated and (b) correlated disorder of different strengths indicated by the values of $k$. With the logarithmic $y$-axis one can observe indeed an ideal linear dependence of the mobility on $\sqrt{F}$ for most of the cases. Exceptions can be noted for $k=1$ with $N_c=1$ in (a) and $k=1$ for both one and two charge carriers in (b), where the shown field-dependence deviates from the Poole--Frenkel model at low $F$. Nevertheless, for all scenarios studied in this section, we have extracted the Poole--Frenkel parameters $\mu_0$ and $\beta$ from the results in Figure~\ref{fig:time_F2} and summarize them in Table~\ref{tab:PFdata}. For $k=0$, there is by construction no difference between the uncorrelated and correlated cases. With an increasing amount of disorder in the system, the mobility increases because the respective ToF increases as discussed in the field-free cases before, which is reflected in the values for $\mu_0$. 

For each value of $k>0$, mobilities for the spatially uncorrelated site-energies are lower than those in the correlated case. From the inverse relation of $\mu$ and $\tau$, this seems at first glance at odds with the ToF for $F=0$ in Figure~\ref{fig:t_RW_KMC}, where for $N_c=2$ shorter $\tau$ are recorded for systems with uncorrelated site-energies than for correlated ones, which would indicate higher mobility for the former. However, the additional drift component of the charge carriers' dynamics in the direction of an applied electric field has strong effects on the ToF that are different for each of the scenarios studied in the work. One can see an indication of this in Figure~\ref{fig:time_F2} comparing the uncorrelated and correlated cases at low fields. In the correlated case, there is the already mentioned deviation from the ideal linear relation between $\mu$ and $\sqrt{F}$ in the Poole--Frenkel plot. This is not visible in the uncorrelated case.

\begin{figure}[tb]
    \centering
    \includegraphics[width=\textwidth]{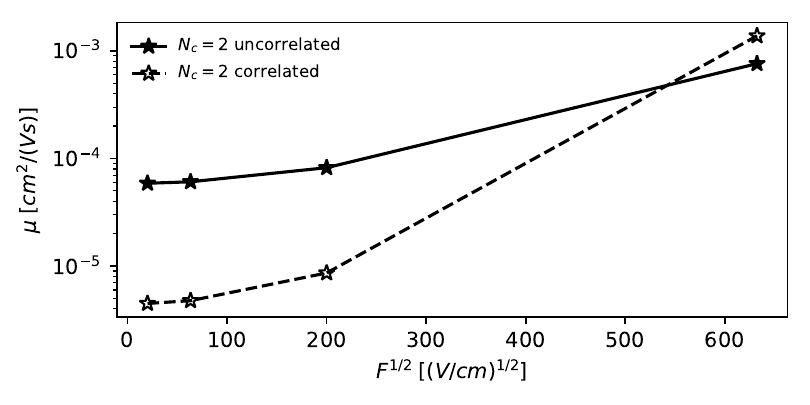}
    \caption{Electric-field dependence of the mobility $\mu$ for small field strengths in the $N=8$ multiscale model system for spatially uncorrelated and spatially correlated disorder for $N_c=2$ and $k=1$, as obtained from the GRW method.}
    \label{fig:smallF}
\end{figure}

We therefore show in Figure~\ref{fig:smallF} the dependence of the mobility on the electric field for much smaller field strengths in the interval $[\unit[4\PT{4}]{V/m},\unit[4\PT{7}]{V/m}]$ for the case of $N_c=2$ and $k=1$. One can clearly see a crossover between the values of the mobility in the uncorrelated and correlated disorder cases for low fields, agreeing qualitatively with what is expected from the ToF data in the diffusive regime discussed in Section~\ref{ssec:GRW_vs_KMC}. This observation of the in general not ideal Poole--Frenkel behavior for small fields emphasizes that the extracted values of $\mu_0$ as listed in Table~\ref{tab:PFdata} should not be interpreted as the true mobilities at $F=0$.

\change{
\section{Scalability and Performance}
\label{sec:scale}
In the previous section, we demonstrated that the GRW method proposed in this work yields quantitatively accurate estimates of the ToF, or the charge mobility in ToF settings, without the need to explicitly calculate the dynamics of charge carriers in multiscale systems and without common sampling and convergence problems of state-of-the art methods, such as KMC. In particular, low charge concentration typically poses the most significant computational challenges for the KMC method, as discussed in Section~\ref{ssec:GRW_vs_KMC} (cf. Figure~\ref{fig:t_step}(b)) and in previous studies~\cite{brereton_efficient_2012,brereton_efficient_2014}. 
Long computational times and inadequate sampling are prevalent issues in most first-principles multiscale systems. As shown in Section \ref{ssec:GRW_vs_KMC} Figure \ref{fig:t_RW_KMC_error}, a 30\% error was observed in our test case with uncorrelated energies and $k=1$. That is, the KMC method reported ToF values that are 30\% larger than the true ToF value. These challenges become more pronounced as the system size and disorder increases. Here, we now turn to a discussion of the scaling and convergence behavior of the GRW method, by considering model systems with different sizes ${n \choose N_c}$ of the state space for different amounts of spatially uncorrelated energetic disorder.
Specifically, we vary the extension of the lattice in the $x$-direction, such that $[0,L_x] \times [0,L] \times [0,L] \subset \mathbb{R}^3 $ where $L=(N-1)a$ with $N=8$ as used in the previous section, and $L_x=(N_x-1)a$ for $N_x=10,20,30,40$. This corresponds to systems with $n=640, 1280, 1920, 2560$, all typical system sizes accessible to first-principle multiscale models. Analogously to the previous section, sites with $x=0$ are source sites, and sites with $x=L_x$ are sink sites, with source and sinks states determined accordingly.

\begin{figure}[tb]
  \centering
  \includegraphics[width=\linewidth]{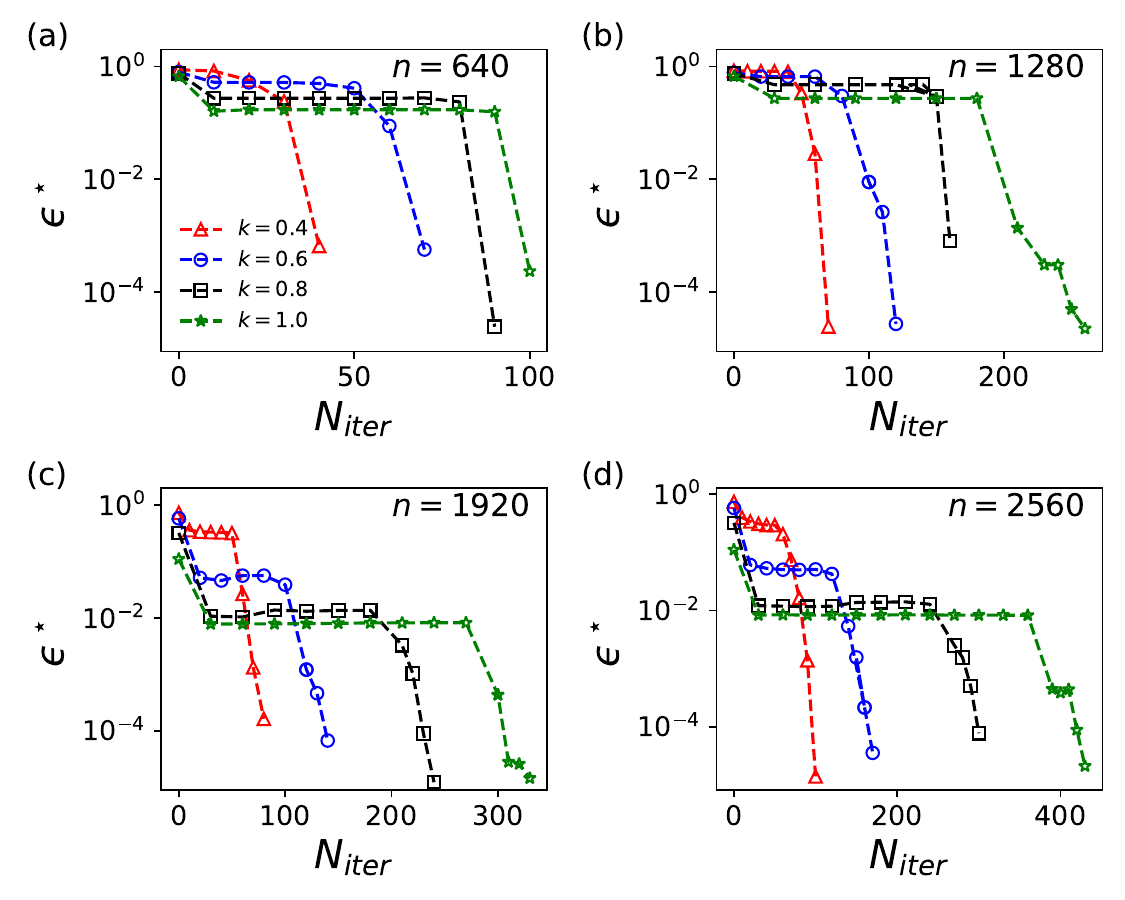}
  \caption{Convergence behavior of the ToF calculation using the GRW method for $N_c=1$. }
  \label{fig:iteration_1c}
\end{figure}

\begin{figure}[tb]
  \centering
  \includegraphics[width=\linewidth]{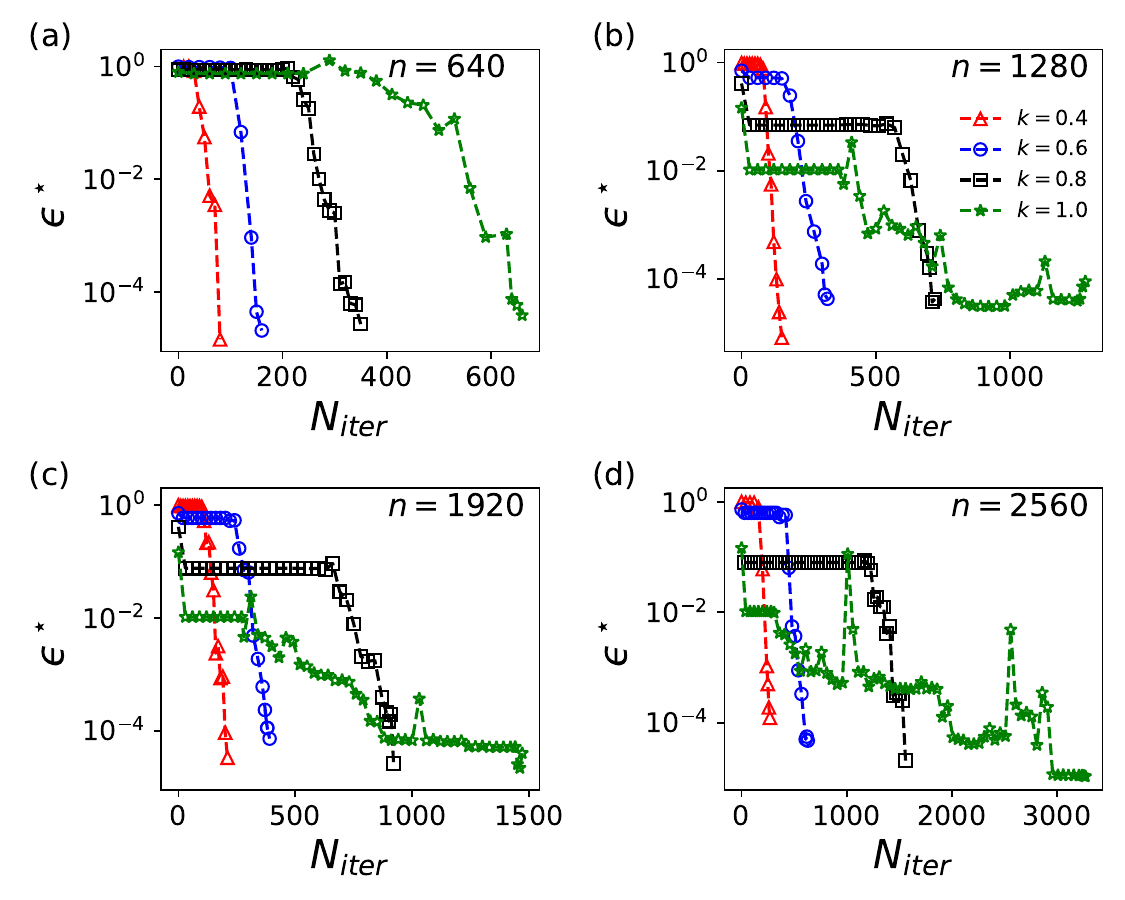}
  \caption{Convergence behavior of the ToF calculation using the GRW method for $N_c=2$. }
  \label{fig:iteration_2c}
\end{figure}

Using the quasi-minimal residual method and the biconjugate gradients stabilized method for solving the matrix Equation~\eqref{equ:matrix1}, the convergence of the solution is quantified using the metric $\epsilon^\star := \| \vec{\tau}^0 \|_2 / \| \vec{\tau}^0 - (\mathbf{I} - \mathbf{P}_1) \vec{\tau}^\ast \|_2$. The convergence of the GRW method is studied by monitoring $\epsilon^\star$ as a function of the number of iterative steps $N_{\text{iter}}$. As described in Section \ref{sec:methods}, convergence is achieved when $\epsilon^\star \leq 10^{-5}$. 
Figure \ref{fig:iteration_1c} illustrates that for $N_c=1$, the number of iterative steps required for convergence increases with both the disorder parameter $k$ and the system size $n$. Specifically, as the system size increases from 640 to 2560, the number of required iterations grows approximately linearly, from 100 to 430.

Figure \ref{fig:iteration_2c} illustrates a similar trend of increasing iteration steps required for convergence. As the system size increases from 640 to 2560, the iteration steps rise substantially from 600 to 3280. The convergence becomes increasingly challenging with higher disorder and larger $n$, as evidenced by the pronounced oscillations. This nonlinear growth arises from the combinatorial increase in states, ${n \choose N_c}$, for $N_c=2$ carriers. In larger systems with $N_c > 2$, the proposed method faces computational challenges due to the exponential growth in state space. At the same time, the KMC method can also have convergence difficulties in large system even if there are multiple carriers.

\begin{figure}[tb]
  \centering
  \includegraphics[width=\linewidth]{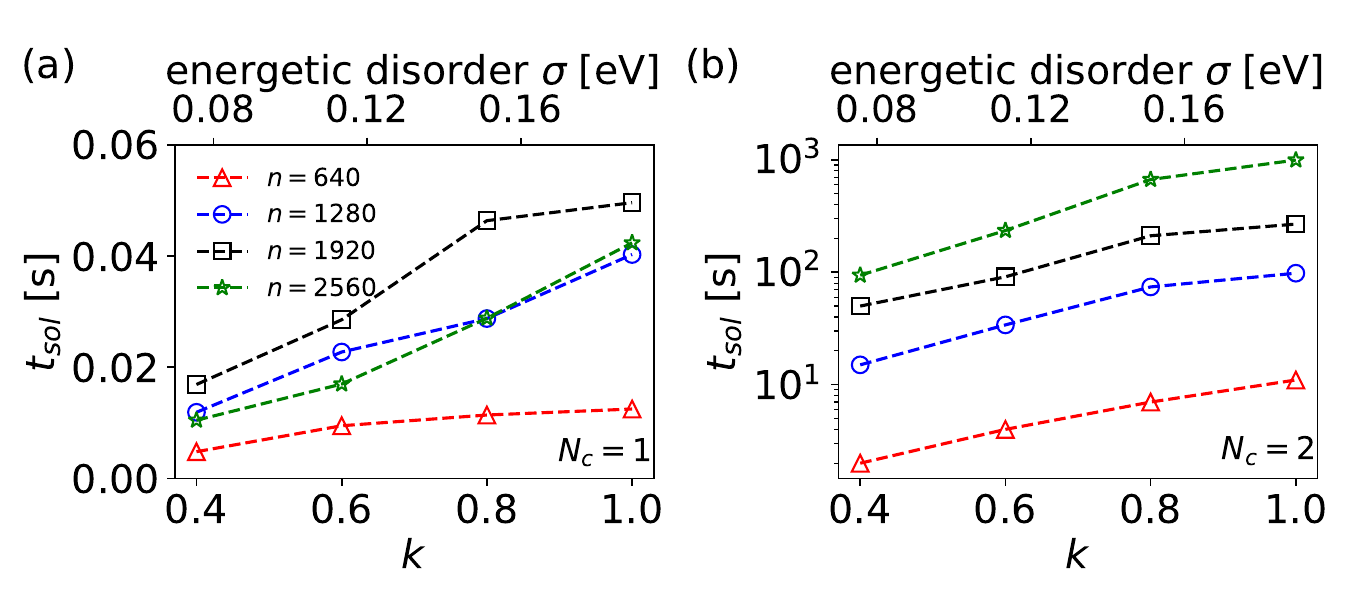}
  \caption{The time-to-solution (in \unit[]{s}) using the GRW method for systems with (a) $N_c=1$ and (b) $N_c=2$ charge carriers and different number of sites depending on the energetic disorder, as recorded with our Python implementation on an Intel(R) Xeon(R) Gold 5120 CPU @ 2.20GHz with 28 threads.}
  \label{fig:timeToSolve}
\end{figure}

From a practical perspective, the time required to obtain the convergence in GRW method, the time-to-solution $t_\text{sol}$, is more relevant than the number of iterations needed. In Figure~\ref{fig:timeToSolve} we show this time for $N_c=1$ and $N_c=2$ and different number of sites with varying energetic disorder, as recorded with our proof-of-concept implementation in Python on an Intel(R) Xeon(R) Gold 5120 CPU @ 2.20GHz with 28 threads. When there is one carrier, the GRW method takes less than \unit[1]{s} to achieve converged ToF, even for the largest system with $n=2560$. In contrast, a single KMC simulation, i.e., for one random seed from one sink state, using production code for the uncorrelated system with size $n=512$ can take more than 1 minute to complete on the same system, and the ToF obtained from 1000 KMC simulations stills has 30\% relative error. Even though the KMC simulations can be trivially parallelized, these aspects highlight the advantages of the GRW method in computational time. When $N_c=2$, one can see that the required computational time increase exponentially. This exponential time increment for multiple carrier systems is a challenge for the GRW method and requires further exploration. 
}

\section{Conclusion}
In this paper, we introduced a Graph Random Walk (GRW) method for calculating the expected time-of-flight of charge carriers in complex molecular materials. A weighted, directed graph model is set up typically from multiscale quantum-classical simulation frameworks and the charge carriers are random walkers. In the case of multiple carriers, the random walk is subject to exclusion to ensure that each vertex of the graph cannot be occupied by more than one charge due to the physical Pauli exclusion. Based on graph theory, the GRW method avoids numerically cumbersome calculations of the explicit dynamics of the carriers, and is therefore not prone to discretization problems or sampling issues as commonly used Master Equation or kinetic Monte Carlo approaches. We have shown that it allows for accurate and reliable predictions of the effective time-of-flight (in the diffusive regime) and the field-dependent charge mobility (in the drift-diffusive regime) for a wide range of scenarios covering vastly different possible material properties. 
\newpage
\appendix

\section{KMC and GRW results data}

\begin{table}[h]
    {\footnotesize
      \caption{Calculated mean ToF $\tau$ and standard error of the mean (in \unit[]{s}) depending on disorder strength on systems with uncorrelated and spatially correlated site energies, obtained from GRW and KMC, respectively. Each data point $\tau$ represents the sample average of the ten realizations of the Gaussian distributed energy landscapes. Each KMC simulation contains 1000 runs.}\label{tab:compare}
    \begin{center}
      \begin{tabular}{cr@{\,$\pm$\,}lr@{\,$\pm$\,}lcr@{\,$\pm$\,}lr@{\,$\pm$\,}l} \hline
        &  \multicolumn{4}{c}{\bf uncorrelated} & &\multicolumn{4}{c}{\bf correlated}\\\cline{2-5}  \cline{7-10}
        $k$  & \multicolumn{2}{c}{\bf GRW} & \multicolumn{2}{c}{\bf KMC} && \multicolumn{2}{c}{\bf GRW} & \multicolumn{2}{c}{\bf KMC}\\\hline
        \multicolumn{10}{c}{$\mathbf{N_c=1}$}\\
    0.0  & (1.13 & 0.00)\PT{-11} & (1.18 & 0.00)\PT{-11} & & (1.13 & 0.00)\PT{-11} & (1.18 & 0.00)\PT{-11}\\
    0.1  & (1.33 & 0.02)\PT{-11} & (1.35 & 0.02)\PT{-11} & & (1.32 & 0.05)\PT{-11} & (1.31 & 0.05)\PT{-11}\\ 
    0.2  & (2.19 & 0.08)\PT{-11} & (2.18 & 0.08)\PT{-11} & & (2.10 & 0.23)\PT{-11} & (2.12 & 0.26)\PT{-11}\\ 
    0.3  & (5.25 & 0.75)\PT{-11} & (5.32 & 0.77)\PT{-11} & & (4.82 & 1.29)\PT{-11} & (4.92 & 1.39)\PT{-11}\\ 
    0.4  & (2.18 & 0.95)\PT{-10} & (2.25 & 1.02)\PT{-10} & & (1.72 & 0.86)\PT{-10} & (1.71 & 0.86)\PT{-10}\\ 
    0.5  & (1.58 & 1.16)\PT{-9}  & (1.68 & 1.26)\PT{-9}  & & (9.37 & 6.65)\PT{-10} & (8.83 & 6.14)\PT{-10}\\ 
    0.6  & (1.39 & 1.21)\PT{-8}  & (1.29 & 1.17)\PT{-8}  & & (6.66 & 5.57)\PT{-9}  & (6.41 & 5.33)\PT{-9}\\ 
    0.7  & (1.22 & 1.12)\PT{-7}  & (1.16 & 1.07)\PT{-7}  & & (5.34 & 4.79)\PT{-8}  & (5.49 & 4.93)\PT{-8}\\ 
    0.8  & (1.04 & 0.97)\PT{-6}  & (1.12 & 1.05)\PT{-6}  & & (4.43 & 4.10)\PT{-7}  & (4.40 & 4.06)\PT{-7}\\ 
    0.9  & (8.82 & 8.32)\PT{-6}  & (8.97 & 8.46)\PT{-6}  & & (3.67 & 3.44)\PT{-6}  & (3.65 & 3.43)\PT{-6}\\ 
    1.0  & (7.34 & 6.95)\PT{-5}  & (9.87 & 9.35)\PT{-5}  & & (2.99 & 2.82)\PT{-5}  & (3.05 & 2.88)\PT{-5}\\[0.2cm]
      \multicolumn{10}{c}{$\mathbf{N_c=2}$}\\
        0.0  & (6.63 & 0.00)\PT{-12} & (6.37 & 0.00)\PT{-12} & & (6.63 & 0.00)\PT{-12} & (6.37 & 0.00)\PT{-12}\\
        0.1  & (7.74 & 0.08)\PT{-12} & (7.76 & 0.08)\PT{-12} & & (7.62 & 0.25)\PT{-12} & (7.47 & 0.25)\PT{-12}\\ 
        0.2  & (1.24 & 0.04)\PT{-11} & (1.23 & 0.04)\PT{-11} & & (1.16 & 0.11)\PT{-11} & (1.16 & 0.11)\PT{-11}\\ 
        0.3  & (2.53 & 0.15)\PT{-11} & (2.53 & 0.15)\PT{-11} & & (2.41 & 0.54)\PT{-11} & (2.43 & 0.55)\PT{-11}\\ 
        0.4  & (5.95 & 0.54)\PT{-11} & (6.00 & 0.56)\PT{-11} & & (6.81 & 2.84)\PT{-11} & (6.89 & 2.88)\PT{-11}\\ 
        0.5  & (1.52 & 0.20)\PT{-10} & (1.51 & 0.20)\PT{-10} & & (2.46 & 1.48)\PT{-10} & (2.43 & 1.43)\PT{-10}\\ 
        0.6  & (4.11 & 0.78)\PT{-10} & (4.11 & 0.81)\PT{-10} & & (1.02 & 0.75)\PT{-9}  & (1.01 & 0.73)\PT{-9}\\ 
        0.7  & (1.17 & 0.31)\PT{-9}  & (1.21 & 0.34)\PT{-9}  & & (4.51 & 3.67)\PT{-9}  & (4.51 & 3.67)\PT{-9}\\ 
        0.8  & (3.44 & 1.21)\PT{-9}  & (3.48 & 1.23)\PT{-9}  & & (2.02 & 1.73)\PT{-8}  & (2.02 & 1.74)\PT{-8}\\ 
        0.9  & (1.05 & 0.46)\PT{-8}  & (1.08 & 0.47)\PT{-8}  & & (8.95 & 7.94)\PT{-8}  & (9.04 & 8.04)\PT{-8}\\ 
        1.0  & (3.30 & 1.71)\PT{-8}  & (3.28 & 1.69)\PT{-8}  & & (3.92 & 3.55)\PT{-7}  & (4.04 & 3.66)\PT{-7}\\ \hline
          \end{tabular}
        \end{center}
        }
        \end{table}\newpage


\begin{thebibliography}{10}

  \bibitem{hughesRandomWalksRandom1996}
  Barry~D. Hughes and Barry~D. Hughes.
  \newblock {\em Random {{Walks}} and {{Random Environments}}: {{Volume}} 2: {{Random Environments}}}.
  \newblock {Oxford University Press}, {Oxford, New York}, June 1996.
  
  \bibitem{Zeitouni2004}
  Ofer Zeitouni.
  \newblock {\em Part II: Random Walks in Random Environment}, pages 190--312.
  \newblock Springer Berlin Heidelberg, Berlin, Heidelberg, 2004.
  
  \bibitem{RIEDE2011448}
  M.~Riede, B.~Lüssem, and K.~Leo.
  \newblock 4.13 - organic semiconductors.
  \newblock In Pallab Bhattacharya, Roberto Fornari, and Hiroshi Kamimura, editors, {\em Comprehensive Semiconductor Science and Technology}, pages 448--507. Elsevier, Amsterdam, 2011.
  
  \bibitem{doi:10.1021/cr040084k}
  Jean-Luc Bredas, David Beljonne, Veaceslav Coropceanu, and J~Cornil.
  \newblock Charge-transfer and energy-transfer processes in conjugated oligomers and polymers: A molecular picture.
  \newblock {\em Chemical Reviews}, 104(11):4971--5004, 2004.
  
  \bibitem{pasveer_unified_2005}
  W.~F. Pasveer, J.~Cottaar, C.~Tanase, R.~Coehoorn, P.~A. Bobbert, P.~W.~M. Blom, D.~M. de Leeuw, and M.~A.~J. Michels.
  \newblock Unified {Description} of {Charge}-{Carrier} {Mobilities} in {Disordered} {Semiconducting} {Polymers}.
  \newblock {\em Physical Review Letters}, 94(20):206601, May 2005.
  
  \bibitem{tof_exp}
  Masahiro Funahashi.
  \newblock {\em Time-of-Flight Method for Determining the Drift Mobility in Organic Semiconductors}, chapter~6, pages 161--178.
  \newblock John Wiley \& Sons, Ltd, 2021.
  
  \bibitem{ruhle_microscopic_2011}
  Victor Rühle, Alexander Lukyanov, Falk May, Manuel Schrader, Thorsten Vehoff, James Kirkpatrick, Björn Baumeier, and Denis Andrienko.
  \newblock Microscopic {Simulations} of {Charge} {Transport} in {Disordered} {Organic} {Semiconductors}.
  \newblock {\em Journal of Chemical Theory and Computation}, 7(10):3335--3345, October 2011.
  
  \bibitem{doi:10.1021/ja305310r}
  Falk May, Mustapha Al-Helwi, Björn Baumeier, Wolfgang Kowalsky, Evelyn Fuchs, Christian Lennartz, and Denis Andrienko.
  \newblock Design rules for charge-transport efficient host materials for phosphorescent organic light-emitting diodes.
  \newblock {\em Journal of the American Chemical Society}, 134(33):13818--13822, 2012.
  
  \bibitem{stenzel_general_2014}
  Ole Stenzel, Christian Hirsch, Tim Brereton, Bjoern Baumeier, Denis Andrienko, Dirk Kroese, and Volker Schmidt.
  \newblock A {General} {Framework} for {Consistent} {Estimation} of {Charge} {Transport} {Properties} via {Random} {Walks} in {Random} {Environments}.
  \newblock {\em Multiscale Modeling \& Simulation}, 12(3):1108--1134, January 2014.
  
  \bibitem{poelkingImpactMesoscaleOrder2015}
  Carl Poelking, Max Tietze, Chris Elschner, Selina Olthof, Dirk Hertel, Bj{\"o}rn Baumeier, Frank W{\"u}rthner, Klaus Meerholz, Karl Leo, and Denis Andrienko.
  \newblock Impact of mesoscale order on open-circuit voltage in organic solar cells.
  \newblock {\em Nature Mater}, 14(4):434--439, April 2015.
  
  \bibitem{omura_master-equation-based_2021}
  Yasuhisa Omura.
  \newblock Master-equation-based approach to stochastic processes in few-electron systems and advanced considerations for practical applications.
  \newblock {\em AIP Advances}, 11(11):115123, November 2021.
  
  \bibitem{kolesnikov_kinetic_2018}
  S.~V. Kolesnikov, A.~M. Saletsky, S.~A. Dokukin, and A.~L. Klavsyuk.
  \newblock Kinetic {Monte} {Carlo} {Method}: {Mathematical} {Foundations} and {Applications} for {Physics} of {Low}-{Dimensional} {Nanostructures}.
  \newblock {\em Mathematical Models and Computer Simulations}, 10(5):564--587, September 2018.
  
  \bibitem{brereton_efficient_2012}
  Tim~J. Brereton, Dirk~P. Kroese, Ole Stenzel, Volker Schmidt, and Bjorn Baumeier.
  \newblock Efficient simulation of charge transport in deep-trap media.
  \newblock In {\em Proceedings {Title}: {Proceedings} of the 2012 {Winter} {Simulation} {Conference} ({WSC})}, pages 1--12, Berlin, Germany, December 2012. IEEE.
  
  \bibitem{baumeier_stochastic_2012}
  Björn Baumeier, Ole Stenzel, Carl Poelking, Denis Andrienko, and Volker Schmidt.
  \newblock Stochastic modeling of molecular charge transport networks.
  \newblock {\em Physical Review B}, 86(18):184202, November 2012.
  
  \bibitem{norris_markov_1998}
  J.~R. Norris.
  \newblock {\em Markov chains}.
  \newblock Cambridge series on statistical and probabilistic mathematics. Cambridge University Press, Cambridge, UK ; New York, 1st pbk. ed edition, 1998.
  
  \bibitem{doyle_random_2000}
  Peter~G. Doyle and J.~Laurie Snell.
  \newblock Random {Walks} and {Electric} {Networks}.
  \newblock 2000.
  \newblock Publisher: arXiv Version Number: 1.
  
  \bibitem{freund_qmr_1991}
  Roland~W. Freund and Noel~M. Nachtigal.
  \newblock {QMR}: a quasi-minimal residual method for non-{Hermitian} linear systems.
  \newblock {\em Numerische Mathematik}, 60(1):315--339, December 1991.
  
  \bibitem{van_der_vorst_bi}
  H.~A. Van Der~Vorst.
  \newblock Bi-{CGSTAB}: {A} {Fast} and {Smoothly} {Converging} {Variant} of {Bi}-{CG} for the {Solution} of {Nonsymmetric} {Linear} {Systems}.
  \newblock {\em SIAM Journal on Scientific and Statistical Computing}, 13(2):631--644, March 1992.
  
  \bibitem{pasveer_scaling_2005}
  W.~F. Pasveer, P.~A. Bobbert, H.~P. Huinink, and M.~A.~J. Michels.
  \newblock Scaling of current distributions in variable-range hopping transport on two- and three-dimensional lattices.
  \newblock {\em Physical Review B}, 72(17):174204, November 2005.
  
  \bibitem{casalegno_numerical_2013}
  Mosè Casalegno, Andrea Bernardi, and Guido Raos.
  \newblock Numerical simulation of photocurrent generation in bilayer organic solar cells: {Comparison} of master equation and kinetic {Monte} {Carlo} approaches.
  \newblock {\em The Journal of Chemical Physics}, 139(2):024706, July 2013.
  
  \bibitem{koster_charge_2010}
  L.~J.~A. Koster.
  \newblock Charge carrier mobility in disordered organic blends for photovoltaics.
  \newblock {\em Physical Review B}, 81(20):205318, May 2010.
  
  \bibitem{yu_molecular_2001}
  Z.~G. Yu, D.~L. Smith, A.~Saxena, R.~L. Martin, and A.~R. Bishop.
  \newblock Molecular geometry fluctuations and field-dependent mobility in conjugated polymers.
  \newblock {\em Physical Review B}, 63(8):085202, February 2001.
  
  \bibitem{jahnke_solving_2008}
  Tobias Jahnke and Steffen Galan.
  \newblock Solving {Chemical} {Master} {Equations} by an {Adaptive} {Wavelet} {Method}.
  \newblock pages 290--293, Psalidi, Kos (Greece), 2008.
  
  \bibitem{wolf_solving_2010}
  Verena Wolf, Rushil Goel, Maria Mateescu, and Thomas~A Henzinger.
  \newblock Solving the chemical master equation using sliding windows.
  \newblock {\em BMC Systems Biology}, 4(1):42, December 2010.
  
  \bibitem{cottaar_modeling_2012}
  J.~Cottaar, R.~Coehoorn, and P.A. Bobbert.
  \newblock Modeling of charge transport across disordered organic heterojunctions.
  \newblock {\em Organic Electronics}, 13(4):667--672, April 2012.
  
  \bibitem{marsh_microscopic_2007}
  R.~A. Marsh, C.~Groves, and N.~C. Greenham.
  \newblock A microscopic model for the behavior of nanostructured organic photovoltaic devices.
  \newblock {\em Journal of Applied Physics}, 101(8):083509, April 2007.
  
  \bibitem{wolf_current_1999}
  U.~Wolf, V.~I. Arkhipov, and H.~Bässler.
  \newblock Current injection from a metal to a disordered hopping system. {I}. {Monte} {Carlo} simulation.
  \newblock {\em Physical Review B}, 59(11):7507--7513, March 1999.
  
  \bibitem{casalegno_coarse-grained_2012}
  Mosè Casalegno, Chiara Carbonera, Silvia Luzzati, and Guido Raos.
  \newblock Coarse-grained kinetic modelling of bilayer heterojunction organic solar cells.
  \newblock {\em Organic Electronics}, 13(5):750--761, May 2012.
  
  \bibitem{FRM}
  Daniel~T. Gillespie.
  \newblock Exact stochastic simulation of coupled chemical reactions.
  \newblock {\em The Journal of Physical Chemistry}, 81(25):2340--2361, 1977.
  
  \bibitem{cheung_modelling_2008}
  David~L. Cheung and Alessandro Troisi.
  \newblock Modelling charge transport in organic semiconductors: from quantum dynamics to soft matter.
  \newblock {\em Physical Chemistry Chemical Physics}, 10(39):5941, 2008.
  
  \bibitem{sato_multiscale_2019}
  Masahiro Sato, Akiko Kumada, and Kunihiko Hidaka.
  \newblock Multiscale modeling of charge transfer in polymers with flexible backbones.
  \newblock {\em Physical Chemistry Chemical Physics}, 21(4):1812--1819, 2019.
  
  \bibitem{brereton_efficient_2014}
  Tim Brereton, Ole Stenzel, Björn Baumeier, Denis Andrienko, Volker Schmidt, and Dirk Kroese.
  \newblock Efficient {Simulation} of {Markov} {Chains} {Using} {Segmentation}.
  \newblock {\em Methodology and Computing in Applied Probability}, 16(2):465--484, June 2014.
  
  \bibitem{bouhassoune_carrier_2009}
  M.~Bouhassoune, S.L.M.~Van Mensfoort, P.A. Bobbert, and R.~Coehoorn.
  \newblock Carrier-density and field-dependent charge-carrier mobility in organic semiconductors with correlated {Gaussian} disorder.
  \newblock {\em Organic Electronics}, 10(3):437--445, May 2009.
  
  \bibitem{baranovskii_theoretical_2014}
  S.~D. Baranovskii.
  \newblock Theoretical description of charge transport in disordered organic semiconductors: {Charge} transport in disordered organic semiconductors.
  \newblock {\em physica status solidi (b)}, 251(3):487--525, March 2014.
  
  \bibitem{PhysRev.54.647}
  J.~Frenkel.
  \newblock On pre-breakdown phenomena in insulators and electronic semi-conductors.
  \newblock {\em Phys. Rev.}, 54:647--648, Oct 1938.
  
  \end{thebibliography}
\end{document}